\documentclass[conference]{IEEEtran}

\usepackage{lipsum}
\usepackage{caption,graphicx,newfloat}
\DeclareCaptionType{List}

\usepackage{booktabs}
\usepackage{multirow}
\usepackage[table,xcdraw]{xcolor}

\usepackage{graphicx}
\usepackage{amsmath}
\usepackage{subfig}
\usepackage[table,xcdraw]{xcolor}
\usepackage{rotating}
\usepackage{multirow}
\usepackage{tikz}

\ifCLASSINFOpdf
\else
\fi

\begin{document}
\title{Evaluating Built-in ECC of FPGA on-chip Memories for the Mitigation of Undervolting Faults}

\author{\IEEEauthorblockN{Behzad Salami\IEEEauthorrefmark{1},
Osman S. Unsal\IEEEauthorrefmark{1}, and
Adrian Cristal Kestelman\IEEEauthorrefmark{1}\IEEEauthorrefmark{2}\IEEEauthorrefmark{3}}
\IEEEauthorblockA{\IEEEauthorrefmark{1}Barcelona Supercomputing Center (BSC), Barcelona, Spain.}
\IEEEauthorblockA{\IEEEauthorrefmark{2}Universitat Politècnica de Catalunya (UPC), Barcelona, Spain.}
\IEEEauthorblockA{\IEEEauthorrefmark{3}IIIA - Artificial Intelligence Research Institute CSIC - Spanish National Research Council, Spain.}
Emails: \{behzad.salami, osman.unsal, and adrian.cristal\}@bsc.es
}

\maketitle

\begin{abstract}

Voltage underscaling below the nominal level is an effective solution for improving energy efficiency in digital circuits, \textit{e.g.}, Field Programmable Gate Arrays (FPGAs). However, further undervolting below a safe voltage level and without accompanying frequency scaling leads to timing related faults, potentially undermining the energy savings. Through experimental voltage underscaling studies on commercial FPGAs, %
 we observed that the rate of these faults exponentially increases for on-chip memories, or Block RAMs (BRAMs).   
To mitigate these faults, we evaluated the efficiency of the built-in Error-Correction Code (ECC) and observed that more than 90\% of the faults are correctable and further 7\% are detectable (but not correctable). This efficiency is the result of the single-bit type of these faults, which are then effectively covered by the Single-Error Correction and Double-Error Detection (SECDED) design of the built-in ECC. %
Finally, motivated by the above experimental observations, we evaluated an FPGA-based Neural Network (NN) accelerator under low-voltage operations, while built-in ECC is leveraged to mitigate undervolting faults and thus, prevent NN significant accuracy loss. In consequence, we achieve 40\% of the BRAM power saving through undervolting below the minimum safe voltage level, with a negligible NN accuracy loss, thanks to the substantial fault coverage by the built-in ECC.%

\end{abstract}

\IEEEpeerreviewmaketitle

\section{Introduction}
The power and energy dissipation of digital circuits is directly related to their supply voltage. %
For instance, in recent years, it has been shown that undervolting, \textit{i.e.}, voltage underscaling below the nominal level that is a factory setting by vendors, can substantially improve the energy efficiency of real hardware including \textit{i)} processing devices, \textit{e.g.}, CPUs \cite{vlv}, Graphics Processing Units (GPUs) \cite{Leng2015}, Field Programmable Gate Arrays (FPGAs) \cite{behzad-thesis}, Application-specific Integrated Circuits (ASICs) \cite{dnnengine} and \textit{ii)} memory modules, \textit{e.g.}, Dynamic RAMs (DRAMs) \cite{Chang2017} and Static RAMs (SRAMs) \cite{sram}. Unlike Dynamic Voltage and Frequency Scaling (DVFS) technique \cite{dvfsfpga}, the frequency is not normally scaled down in the undervolting approach. Therefore energy savings can be more significant; however, timing related faults can appear, which in turn, can cause applications to crash or terminate with wrong results. %
Efficiently mitigating these faults can allow further undervolting. However, existing undervolting fault mitigation techniques need either extensive hardware or redesign. For instance, Razor \cite{Ernst2003} that dynamically underscales the voltage until a fault occurs leverages additional delay latches; and \cite{Chang2017} requires major modifications on the memory controller to deal with the reduced supply voltages. Alternatively, we aim to leverage the built-in Error Correction Code (ECC) of FPGAs to mitigate these faults, due to increasing iterest to exploit FPGAs systems in different domains such as query processing \cite{axle1}, \cite{axle2}, \cite{axle3}, \cite{axle4}, Neural Network (NN) \cite{dnn}, among others. ECC is a class of Hamming code-based fault mitigation technique that is conventionally used to mitigate soft errors \cite{softerror}. This paper aims to evaluate its efficiency in aggressive low-voltage regions, to the best of our knowledge for the first time in commercial FPGAs. 

The concentration of this paper is on FPGA-based on-chip memories, or Block RAMs (BRAMs), as a major FPGA component. BRAMs of the studied commercial FPGAs from Xilinx, a main vendor, are equipped with a Single-Error Correction and Double-Error Detection (SECDED) ECC mechanism \cite{bram}. Through experiments on a VC707 and two identical samples of KC705 FPGA platforms, we find that with undervolting below the nominal level and until $V_{min}$, no observable faults occur; however, with further voltage reduction faults manifest with an exponentially increasing rate, after which the system crashes at $V_{crash}$. The conservative voltage margin is confirmed to be on average 39\% of the nominal level, set by the vendor to ensure the worst-case process and environmental conditions. Also, through undervolting below the voltage guardband from $V_{min}$ to $V_{crash}$, BRAMs power consumption is further reduced; however, faults appear in this region require fault mitigation- in our case by leveraging ECC.%

We experimentally find that the built-in ECC does not incur considerable power overhead while showing good coverage of undervolting faults with a 90\% of correction and a further 7\% detection (but not correction) capability. Our experiments indicate that most of the undervolting faults manifest themselves single-bit faults, which can be very efficiently corrected by the SECDED type built-in ECC. Also, we discover that by further undervolting, correctable faults manifest before detectable faults and in turn, detectable faults manifest before undetectable faults. This behavior is the consequence of the Fault Inclusion Property (FIP), \textit{i.e.}, all faulty bits, if any, at a certain voltage level would also be faulty (and expanded to other bits) at lower voltage levels. Due to this observation of faults behavior, application-aware run-time undervolting techniques can be deployed, where the supply voltage is underscaled until the first fault is detected (but not corrected) by the built-in ECC. Finally, motivated by above experimental observations, we evaluate an FPGA-based NN accelerator under low-voltage BRAMs operations, while built-in ECC is leveraged to cover undervolting faults and in turn, to prevent accuracy loss. In consequence, the total power consumption of the accelerator is reduced by 25.2\% through undervolting BRAMs from the nominal level to the $V_{crash}$, with a negligible NN accuracy loss. 

\textbf{{Overall, this paper aims to evaluate the built-in ECC of aggressively-undervolted BRAMs of commercial FPGAs}}, associated with the following contributions:      

\begin{itemize}
\item Evaluating the trade-off among fault coverage capability and power consumption, in FPGA BRAMs equipped with built-in ECC at low-voltage regions.
\item Characterizing the behavior of faults according to the inherent capability of the built-in ECC, \textit{i.e.}, SECDEC under aggressive low-voltage operations.
\item Evaluating a low-voltage FPGA-based NN accelerator, protected by built-in ECC to prevent accuracy loss.
\end{itemize}

This paper is organized as follows. In Section II, we discuss the FPGA BRAMs undervolting methodology and experimental observations. In Section III, we elaborate on the efficiency of the built-in ECC for the fault mitigation, with a case study evaluation on NN accelerators, as discussed in Section IV. Section V reviews the previous work, and finally, Section VI concludes and summarizes the paper.
\section{Understanding FPGA BRAMs Undervolting }

We perform our undervolting experiments on two representative Xilinx FPGA platforms, \textit{i.e.}, VC707 and KC705, representing performance- and power-optimized designs, respectively. Also, for investigating the impact of the die-to-die process variation, we repeat our experiments on two identical samples of KC705. VC707 and KC705 are equipped with 2060 and 890 BRAMs, respectively, distributed all over the chip with the size of 18-kbits each and with a nominal voltage of $V_{nom}= 1V$. Each BRAM is a matrix of bitcells with 1024 rows and 18 columns. When ECC is activated, 2-bits per row are reserved as parity; otherwise, all 18-bits can be used as data bits. For the flexibility purpose, BRAMs can be either individually accessed or cascaded to build larger memories. %
Note thatfor the voltage scaling of BRAMs ($V_{CCBRAM}$) we use the same methodology of \cite{behzad-micro}, which is based on Power Management Bus (PMBus) \cite{pmbus}. %

\begin{figure*}[!t]
\centering
\subfloat[VC707]{\includegraphics[width=0.33\linewidth]{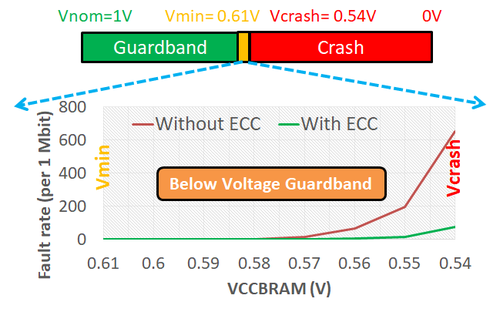}%
\label{subfig:eccmajor707}}
\hfill
\subfloat[KC705-A]{\includegraphics[width=0.33\linewidth]{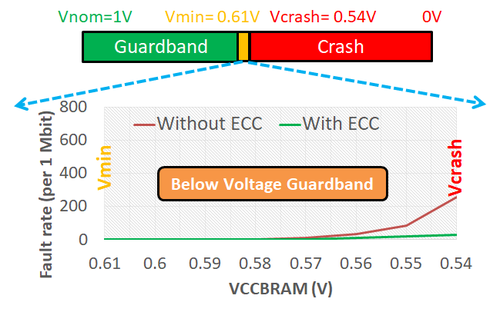}%
\label{subfig:eccmajor705A}}
\hfill
\subfloat[KC705-B]{\includegraphics[width=0.33\linewidth]{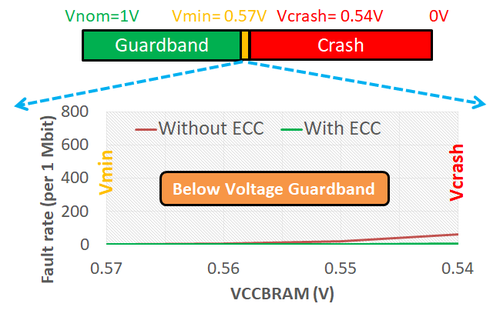}%
\label{subfig:eccmajor705B}}
\caption{The efficiency of ECC to mitigate undervolting faults at the critical voltage regions below $V_{min}$.}
\label{fig:eccmajor}
\end{figure*}

 Finally, to measure the power consumption, we use real time power measurement by the voltage regulator reported by PMBus. 
 We report the total power consumption savings, including dynamic and static parts through the undervolting. Note that experiments are performed on the default and fixed internal frequency of BRAMs, {i.e.}, $\sim$500MHz \cite{bram}. In short, we gradually undervolt $V_{CCBRAM}$ until system crashes, while reading contents of BRAMs. For each voltage level, we record the fault rate, fault location, and power consumption of BRAMs for further analysis. %
\subsection{BRAM Configuration} 
In studied platforms, there are several options for BRAMs configurations. Our experimental setup is based on the following configurations:
\begin{itemize}
\item Configuration modes: We use simple dual-port mode BRAMs since it is the only mode that ECC can be activated. However, the overall voltage behavior (large voltage guardband, critical, and crash voltage regions) has experimented on other BRAM modes, and quite similar behavior is observed. 
\item Soft- vs. hard-core ECC: Two types of ECC are available in Xilinx BRAMs, \textit{i.e.}, soft- and hard-core with the same functionality. Unlike the hard-core, in the soft-core ECC, FPGA resources such as LUTs are utilized to implement the corresponding functionality. Thus, we make our study on hard-core built-in ECC, which does not require any additional hardware.
\item Word width: Our design is based on memory width of 64-bits since the built-in ECC is optimized for memories with word-width $\ge$ 64-bits \cite{bram}. Note that since the basic BRAMs word-width is 18-bits, the memory used in our study is built by automatically cascading original BRAMs. 
\end{itemize}

Finally, note that we access BRAMs in the non-buffered mode without exploiting RAM FIFO and output registers; thus, any read operation on-the-fly and directly accesses to BRAMs; in other words, buffered data is not accessed.

\subsection{Overall Power and Fault Rate Behavior} 
We experimentally observe that by undervolting below the nominal voltage level, \textit{i.e.}, $V_{nom}= 1V$ two voltage thresholds exist. \textit{First}, minimum safe voltage, \textit{i.e.}, $V_{min}$, which separates the fault-free or voltage guardband margin and faulty regions. This voltage guardband is confirmed to be on average \%39 of the nominal voltage for studied platforms under ambient temperature and normal environmental conditions; although, slightly different among platforms. This large gap is added by the vendor to ensure the worst-case scenarios. In our experiments, eliminating it delivers more than an order of magnitude power savings without compromising reliability or performance. The voltage guardband gap is also confirmed for other devices such as 12\% for CPUs \cite{itanium}, 20\% for GPUs \cite{Leng2015}, and 16\% for DRAMs \cite{Chang2017}. \textit{Second}, $V_{crash}$ that is the lowest voltage level that our design practically operates, below that FPGA fails. With a slight difference among the FPGA platforms that we studied, $V_{crash}$ is experimentally measured to be 0.54V. %

When $V_{CCBRAM} >= V_{min}$, no observable faults occur; however, %
further undervolting below $V_{min}$, the fault rate exponentially increases, for instance in VC707, up to 0.06\% or 652 faults per 1Mbit %
while the power consumption is further reduced, by 36.1\%. As can be seen in Fig. \ref{fig:eccmajor}, the fault rate in VC707 is significantly higher than KC705, which can be the consequence of their architectural differences. %
Also, a significant difference is observed between the two samples of KC705. As can be seen, KC705-A shows a 4.1$\times$ higher fault rate than KC705-B, which can be due to the die-to-die process variation. This experimental comparison concludes that reliability degradation through aggressive voltage underscaling not only depends on the architecture of the underlying FPGA but also, it can significantly vary for different FPGAs of the same platform. Also, by leveraging ECC in BRAMs, the fault rate in the critical voltage region below $V_{min}$ is significantly reduced by an average of more than 90\% for all platforms, as summarized in Fig. \ref{fig:eccmajor}, and elaborated in Section III.

\section{Efficiency of FPGA BRAMs Built-in ECC under Low-voltage Operations}
The built-in ECC mechanism of BRAMs uses Hamming code. When ECC is activated, parity bits are generated during each write operation and stored along with the data, at the granularity of a single row/word. These parity bits are used during each read operation of a row to correct single-bit faults, or to detect (but not correct) any double-bit fault, termed SECDED. This section evaluates the efficiency and overhead of this ECC in aggressive low-voltage FPGA BRAMs, according to the behavior of undervolting faults by undervolting $V_{CCBRAM}$ below $V_{min}$. Note that since the experimental results on the studied platforms lead to very similar conclusions, in this section and to save the space, we present experimental results on the only VC707.

\begin{figure}[!t]
\centering
\subfloat[Illustration of different fault types in row-column format basic-size BRAM.]{\includegraphics[width=0.9\linewidth]{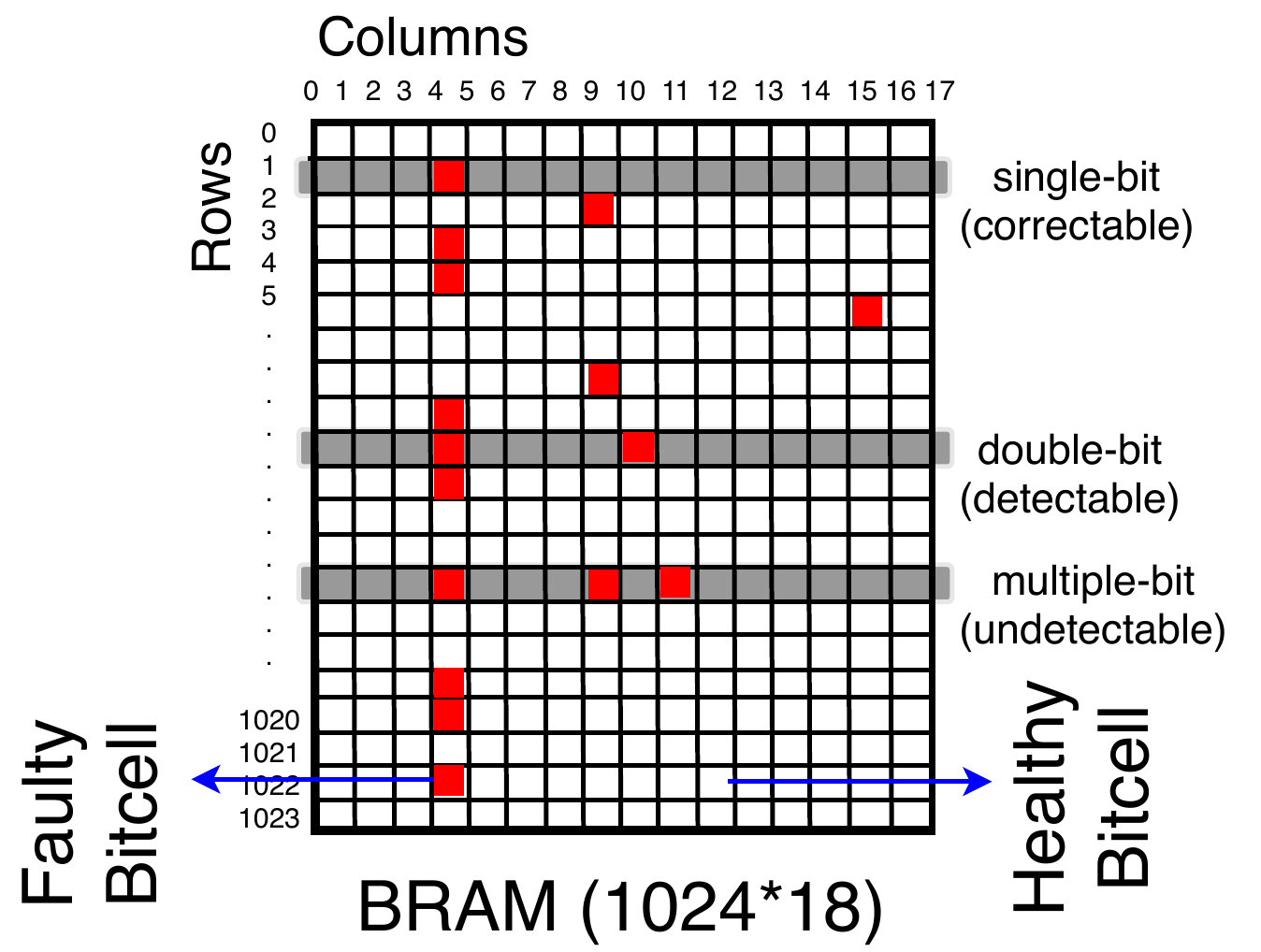}%
\label{subfig:bram1}}
\hfill
\subfloat[Histogram of different fault types under different voltage levels.]{\includegraphics[width=\linewidth]{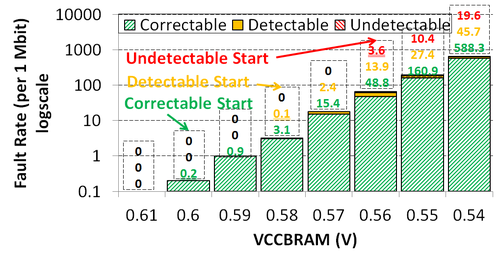}%
\label{subfig:faults}}
\hfill
\subfloat[Fault Inclusion Property (FIP): Faulty bitcells stay faulty at lower voltage levels.]{\includegraphics[width=\linewidth]{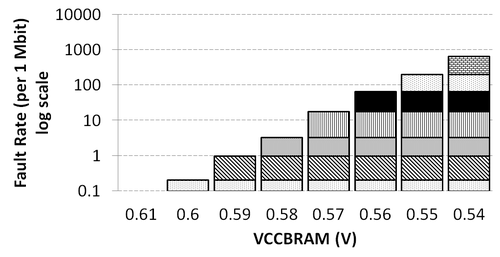}%
\label{subfig:variability1}}
\caption{The Behavior of ECC-activated BRAMs faults, when $V_{CCBRAM}$ is scaled from $V_{min}=0.61V$ to $V_{crash}=0.54V$.}
\label{fig:fip}
\end{figure}

\subsection{Built-in ECC Capability} 
Due to the capability of the built-in ECC in FPGA BRAMs, we categorize faults into correctable (or single-bit), detectable (or double-bit), and undetectable (or multiple-bit) faults, as illustrated in Fig. \ref{subfig:bram1}. Fig. \ref{subfig:faults} shows a histogram of these fault types, in different voltage levels at the critical voltage region, \textit{i.e.}, from $V_{min}=0.61V$ to $V_{crash}=0.54V$ on VC707. We observe that: 

\begin{itemize}
\item The vast majority of these faults are correctable or detectable (but not correctable) by the built-in ECC; for instance, more than 90\% and a further 7\% at $V_{crash}=0.54V$ are correctable and detectable, respectively, using the built-in ECC. This efficiency is the consequence of the inherent type of the built-in ECC, \textit{i.e.}, SECDED, which we experimentally find that it has very good fault coverage due to the relatively sparse distribution of undervolting faults. 
\item By further voltage underscaling, correctable faults manifest before detectable, and in turn, detectable faults manifest before undetectable faults. Through this observation, we leverage the built-in ECC to discover the minimum safe voltage of FPGA-based NN accelerator, as discussed in Section IV.%

The faults behavior mentioned above is the consequence of the Fault Inclusion Property (FIP). FIP is said to exist if all the faulty bits in a certain level of $V_{CCBRAM}$ are still faulty in further reduced voltage levels. In other words, within a memory row which experiences faults, by further undervolting, those initial faulty bits are still faulty and also potentially expanded to other bits. Consequently, single-bit faults can be potentially converted to double-bit and similarly, double-bit faults can be potentially converted to multiple-bit faults. FIP was previously observed for CPU cache structures \cite{Gottscho2014}, here we confirm that FIP holds for FPGA BRAMs, as well, as shown as a stacked chart in Fig. \ref{subfig:variability1}. Also, as noted, FIP results in the behavior as mentioned earlier of the appearance of correctable/detectable/undetectable faults in ECC-activated FPGA BRAMs. %
\end{itemize}

\begin{table}
\centering
\caption{Power and area overheads of the built-in ECC.}
\label{table-eccOverhead}
\begin{tabular}{llll}
\multicolumn{1}{c}{\textbf{}}     & \multicolumn{3}{c}{\textbf{a) Area Utilization (\%)}}                                                 \\ \cline{2-4} 
\multicolumn{1}{l|}{}             & \multicolumn{1}{c|}{BRAM}     & \multicolumn{1}{c|}{LUT}         & \multicolumn{1}{c|}{FF}            \\ \hline
\multicolumn{1}{|l|}{Without ECC} & \multicolumn{1}{r|}{96\%}     & \multicolumn{1}{r|}{3\%}         & \multicolumn{1}{r|}{0.25\%}        \\ \cline{1-1}
\multicolumn{1}{|l|}{With ECC}    & \multicolumn{1}{r|}{100\%}    & \multicolumn{1}{r|}{12\%}        & \multicolumn{1}{r|}{0.25\%}        \\ \hline
\multicolumn{1}{c}{\textbf{}}     & \multicolumn{3}{c}{\textbf{b) BRAM Power (W)}}                                                        \\ \cline{2-4} 
\multicolumn{1}{l|}{}             & \multicolumn{1}{c|}{Vnom= 1V} & \multicolumn{1}{c|}{Vmin= 0.61V} & \multicolumn{1}{c|}{Vcrash= 0.54V} \\ \hline
\multicolumn{1}{|l|}{Without ECC} & \multicolumn{1}{r|}{2.4}      & \multicolumn{1}{r|}{0.31}        & \multicolumn{1}{r|}{0.198}         \\ \cline{1-1}
\multicolumn{1}{|l|}{With ECC}    & \multicolumn{1}{r|}{---}      & \multicolumn{1}{r|}{---}         & \multicolumn{1}{r|}{0.211}         \\ \hline
\multicolumn{4}{l}{---: Above $V_{min}$, since there is no fault, no need for the ECC.}                                                        \\
\multicolumn{4}{c}{Tested Memory Size: 512 * (1024 * 64-bits)}                                                                                 
\end{tabular}
\end{table}

\subsection{Built-in ECC Overhead}

TABLE \ref{table-eccOverhead}(a) includes the area utilization rate of the hardware design described in Section II.A, to evaluate the area cost of the built-in ECC. Toward this goal, our hardware design accesses 512 memories each with the size of 1024 rows of 64-bits, which leads to a full BRAMs utilization on VC707. As can be seen, the built-in ECC does not incur considerable area cost since it is a hard-core unit and internally embedded within the BRAMs structure.    
 Also, TABLE \ref{table-eccOverhead}(b) includes the power overhead of the built-in ECC. We report the power consumption of BRAMs at $V_{nom}=1V, V_{min}=0.61V$, and $V_{crash}=0.54V$. As can be seen, the ECC power overhead is 13mW or 4.2\% at $V_{crash}= 0.54V$. In other words, the power consumption of BRAMs are reduced from 0.31W to 0.211W (31.9\% power reduction) with the voltage underscaling from $V_{min}= 0.61V$ to $V_{crash}= 0.54V$, by exploiting built-in ECC to cover a vast majority of faults. %

\section{The Efficiency of BRAMs ECC for Low-voltage FPGA-based NN Accelerator}
In this section, we discuss the efficiency of the built-in ECC of BRAMs on an FPGA-based NN accelerator, as a case study, under low-voltage operations at the critical voltage region. We follow the same implementation methodology of \cite{behzad-micro} to map NN on FPGA. In this section, we present the experimental results on VC707 for MNIST dataset \cite{mnist}; however, we confirm the conclusions for other studied FPGA platforms as well as for other NN benchmarks such as Forest \cite{forest} and Reutres \cite{reuters}.

\begin{figure}
  \includegraphics[width=\linewidth]{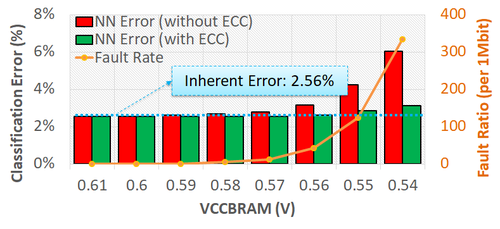}
  \caption{Impact of undervolting faults on the classification error of FPGA-based NN and the efficiency of built-in ECC. $V_{CCBRAM}$ is underscaled at the critical region, \textit{i.e.,} from $V_{min}=0.61V$ to $V_{crash}= 0.54V$ for VC707.}
  \label{fig:dnncharacterization}
\end{figure}

Further voltage underscaling below the voltage guardband and toward $V_{crash}=0.54V$, reduces a further 40\% of BRAMs power over $V_{min}=0.61V$.%
 However, due to undervolting faults in some of BRAMs bitcells, the NN accuracy is degraded. In turn, the classification error is increased from 2.56\% (fault-free classification error at the nominal voltage level) to 6.15\% when $V_{CCBRAM}= V_{crash}= 0.54V$, see Fig. \ref{fig:dnncharacterization}. The NN classification error (left y-axis) increases exponentially, correlated directly with the fault rate increase in BRAMs (right y-axis), as expected. %

To attain the power savings gain of the accelerator without compromising the NN accuracy, we leverage built-in ECC of FPGA BRAMs. In consequence, the NN classification error rate substantially reduces, thanks to the significant fault coverage by ECC, as shown in Fig. \ref{fig:dnncharacterization}.  %
 For instance, the NN classification error has a 0.56\% overhead, \textit{i.e.}, the NN classification error of 2.56\% as the fault-free classification at the nominal voltage level increases to 3.12\% at $V_{CCBRAM}=V_{crash}=0.54V$, when BRAMs are equipped with built-in ECC. This overhead is 6.1$\times$ less than experiments on default BRAMs configuration without ECC, \textit{i.e.}, 3.44\% vs. 0.56\%. %

\section{Related Work}

As an effective solution to reduce power and energy consumption, undervolting is recently studied in detail for modern processors \cite{vlv}, \cite{itanium}, \cite{multicore}, \cite{alperProcessor}, \cite{moons}. Also, GPUs \cite{Leng2015}, DRAMs \cite{Chang2017}, and SRAMs \cite{sram} are evaluated in low-voltage regimes. In parallel, FPGAs have also recently been experimented in part under low-voltage operations \cite{behzad-fpl1}, \cite{behzad-fpl2}, \cite{behzad-micro}, \cite{intel-fpl}; however, without proposing effective fault mitigation techniques. Industrial-based projects have also focused on improving the energy efficiency of hetregenous systems through undervolting \cite{legato1}, \cite{legato2}. All these studies indicate that aggressive undervolting below the voltage guardband margin leads to timing related faults generation. To mitigate these faults, several solutions are investigated, as briefly listed below:
\begin{itemize}
\item Frequency underscaling: By accompanying frequency underscaling, faults can disappear; however, energy efficiency improvements can be limited \cite{intel-fpl}. Also, the implementation cost of this approach, \textit{i.e.}, DVFS is the need for additional online delay monitoring dedicated hardware. Razor is an example of such a hardware mechanism that employs shadow timing latches \cite{Ernst2003}. Recent FPGA implementations of this approach shows 70\% energy saving \cite{Levine14}, \cite{dvfsfpga}, \cite{Nunez2017}. Alternatively, our approach more aggressively allows FPGAs to experience faults with undervolting and later on, employs the built-in ECC fault mitigation technique of the underlying hardware, \textit{i.e.}, FPGA BRAMs.
\item Voltage-adapted device controller: \cite{Chang2017} extensively investigated the effect of undervolting in DRAMs and to mitigate faults, a modified version of the memory controller is presented. The cost of this approach is heavy memory controller modifications. 
\end{itemize} 

Alternatively, we exploit the built-in ECC of BRAMs to cover undervolting faults. This approach does not require any software or hardware modification of the available FPGA design. To the best of our knowledge, such a thorough study on the built-in fault mitigation techniques in ultra low-voltage regions has not been undertaken for FPGAs; although, it is in part studied for CPUs \cite{itanium} and caches \cite{customecc}. %

Motivated by the efficiency of built-in ECC, we exploit it to prevent the accuracy loss in an ultra-low-voltage FPGA-based NN accelerator. To the best of our knowledge, there is not much publicly-available study on NNs with the voltage scaling of the underlying hardware. Existing works are either based on simulations \cite{minerva}, \cite{ThUnderVolt}, \cite{sung2018}, \cite{behzad-hpml} or are partial undervolting studies for customized circuits, \textit{i.e.}, SRAMs \cite{sram} and ASICs \cite{dnnengine}.%

\section{Conclusion and Future Work}
This paper experimentally evaluated the efficiency of the built-in ECC of FPGA BRAMs under aggressive low-voltage operations. Experimental results show the significant efficiency of the built-in ECC, \textit{i.e.}, fault mitigation without considerable power cost. Motivated by these results, we leveraged the built-in ECC to prevent the accuracy loss in low-power FPGA-based Neural Network (NN) accelerator under aggressive low-voltage operations. In consequence, below the safe voltage guardband level, we achieve 40\% power saving with the cost of 0.06\% of the NN accuracy loss. 
As an ongoing work, we are working on the design of customized ECC designs, accounting to our observations of the fault behavior in extremely low-voltage regions of FPGAs, and applying them to the real-world applications.

\section*{Acknowledgment}
The research leading to these results has received funding from the European Union's Horizon 2020 Programme under the LEGaTO Project (www.legato-project.eu), grant agreement n$^\circ$ 780681.

\end{document}